# Machine-learning-assisted analysis of transition metal dichalcogenide thin-film growth

Hyuk Jin Kim[1,#], Minsu Chong[1,#], Tae Gyu Rhee[1,2], Yeong Gwang Khim[1,2], Min-Hyoung Jung[3], Young-Min Kim[3], Hu Young Jeong[4], Byoung Ki Choi[1,5], Young Jun Chang[1,2,*]

[1] Department of Physics, University of Seoul, Seoul, 02504, Republic of Korea,

[2] Department of Smart Cities, University of Seoul, Seoul, 02504, Republic of Korea,

[3] Department of Energy Science, Sungkyunkwan University (SKKU), Suwon, 16419, Republic of Korea,

[4] Graduate School of Semiconductor Materials and Devices Engineering, Ulsan National Institute of Science and Technology (UNIST), Ulsan, 44919, Republic of Korea,

[5] Advanced Light Source (ALS), E. O. Lawrence Berkeley National Laboratory, Berkeley, California 94720, USA

[#]Hyuk Jin Kim and Minsu Chong contributed equally to this work

*Corresponding author: Young Jun Chang

Tel: +82-2-6490-2654

E-mail: yjchang@uos.ac.kr

## Abstract

*In situ* reflective high-energy electron diffraction (RHEED) is widely used to monitor the surface crystalline state during thin-film growth by molecular beam epitaxy (MBE) and pulsed laser deposition. With the recent development of machine learning (ML), ML-assisted analysis of RHEED videos aids in interpreting the complete RHEED data of oxide thin films. The quantitative analysis of RHEED data allows us to characterize and categorize the growth modes step by step, and extract hidden knowledge of the epitaxial film growth process. In this study, we employed the ML-assisted RHEED analysis method to investigate the growth of 2D thin films of transition metal dichalcogenides (ReSe$_2$) on graphene substrates by MBE. Principal component analysis (PCA) and K-means clustering were used to separate statistically important patterns and visualize the trend of pattern evolution without any notable loss of information. Using the modified PCA, we could monitor the diffraction intensity of solely the ReSe$_2$ layers by filtering out the substrate contribution. These findings demonstrate that ML analysis can be successfully employed to examine and understand the film-growth dynamics of 2D materials. Further, the ML-based method can pave the way for the development of advanced real-time monitoring and autonomous material synthesis techniques.

## 1. Introduction

Advanced thin-film synthesis methods, such as molecular beam epitaxy (MBE), pulsed laser deposition (PLD), and atomic layer deposition (ALD), have allowed the formation of atomically sharp interfaces and precise surface engineering in transition metal oxides, III-V semiconductors, and two-dimensional (2D) transition metal dichalcogenides (TMDCs) [1–4]. *In situ* monitoring techniques, such as reflection high-energy electron diffraction (RHEED), spectroscopic ellipsometry, and Auger electron spectroscopy, enable us to monitor the physical properties during the film growth in real time [5–7]. Such *in situ* monitoring techniques have drastically improved our understanding of the growth dynamics. Notably, *in situ* RHEED, which involves the use of high-energy electrons along the grazing incident angle, is sensitive to the topmost surface. Its image data carry a wealth of physical information, such as surface crystallinity, surface morphology, growth rate, in-plane lattice spacing, strain effect, degree of disorder, and changes in surface reconstruction [8–11]. Although the advanced RHEED technique is widely used for the growth of thin films as well as nanostructures, such as nanodots and nanorods [12], only a small fraction of the RHEED data is used. This minute fraction contains static diffraction patterns obtained at a specific time or intensity profile from several diffraction points during the thin-film growth.

With the development of artificial intelligence technology, one should consider adopting machine learning (ML) methods for analyzing the complete RHEED data to advance the existing thin-film growth methods and design fully autonomous material synthesis techniques [13–16]. Deep learning models, such as convolutional neural networks, classified the surface pattern and reconstruction of GaAs [17] and $Fe_xO_y$ [18] with a high accuracy based on the RHEED data. The surface evolution and transitions in an entire RHEED data sequence were also examined for various oxide materials using unsupervised ML methods such as principal component analysis (PCA) and K-means clustering [19–21]. They are advantageous for distinguishing the film-growth dynamics and investigating the time-dependent growth mechanisms and transitions of surface crystalline phases. PCA is an orthogonal linear transformation that defines new orthonormal basis vectors called principal components. Each principal component corresponds to an extracted pattern with a statistical significance (Fig. 1(b)). For the oxide film growth, PCA facilitates the identification of growth modes and reduction of data dimensionality [19,20]. K-means clustering is a vector quantization method in which the RHEED image sequence is partitioned into $K$ clusters based on statistical similarity (Fig. 1(c)). This method allows the identification of stoichiometric changes, strain relaxation, surface reconstruction, and growth mode transitions [19,21].

The ML-assisted RHEED analysis has been applied to analyze the film growth of many oxide materials [18–21], but not for 2D materials. Understanding the growth mechanisms of ultrathin 2D TMDCs is vital for investigating the unique physical properties arising from their 2D van der Waals layered structures. The film growth mechanism of 2D materials is significantly different from that of other oxides, whose interlayer bonding at the interfaces is strong. Typically, 2D materials can grow epitaxially even for a large lattice mismatch between the film and the substrate, because of their weak van der Waals bonding at the interfaces [1]. The growth mechanism of 2D materials has been investigated using *ex situ* characterizations, such as Raman spectroscopy, photoelectron spectroscopy, scanning tunneling microscopy, and transmission electron microscopy [22–25]. These *ex situ* approaches provide limited information on the real-time film growth dynamics, and thus, it is imperative to adopt a suitable method for investigating the entire RHEED video of the film growth of 2D materials.

In this study, we demonstrate the ML-assisted RHEED analysis of TMDC thin-film growth based on unsupervised ML approaches, including PCA and K-means clustering. Using these methods, we can isolate the RHEED patterns based on their statistical importance and then separately monitor the film contributions. The ML-assisted RHEED analysis was primarily conducted on 1T'-$ReSe_2$ thin films grown on graphene substrates by MBE. We developed a modified version of the PCA to detect the thickness oscillation of the 2D thin films by eliminating the strong substate contributions and by reconstructing the RHEED intensity profile of only the thin films. Furthermore, compression of the first thickness oscillation suggested an abrupt change in the film growth rate during the initial growth period. These findings reveal that implementing ML analysis is suitable for attaining a deeper understanding of the film-growth dynamics of 2D materials and for developing advanced real-time film monitoring techniques.

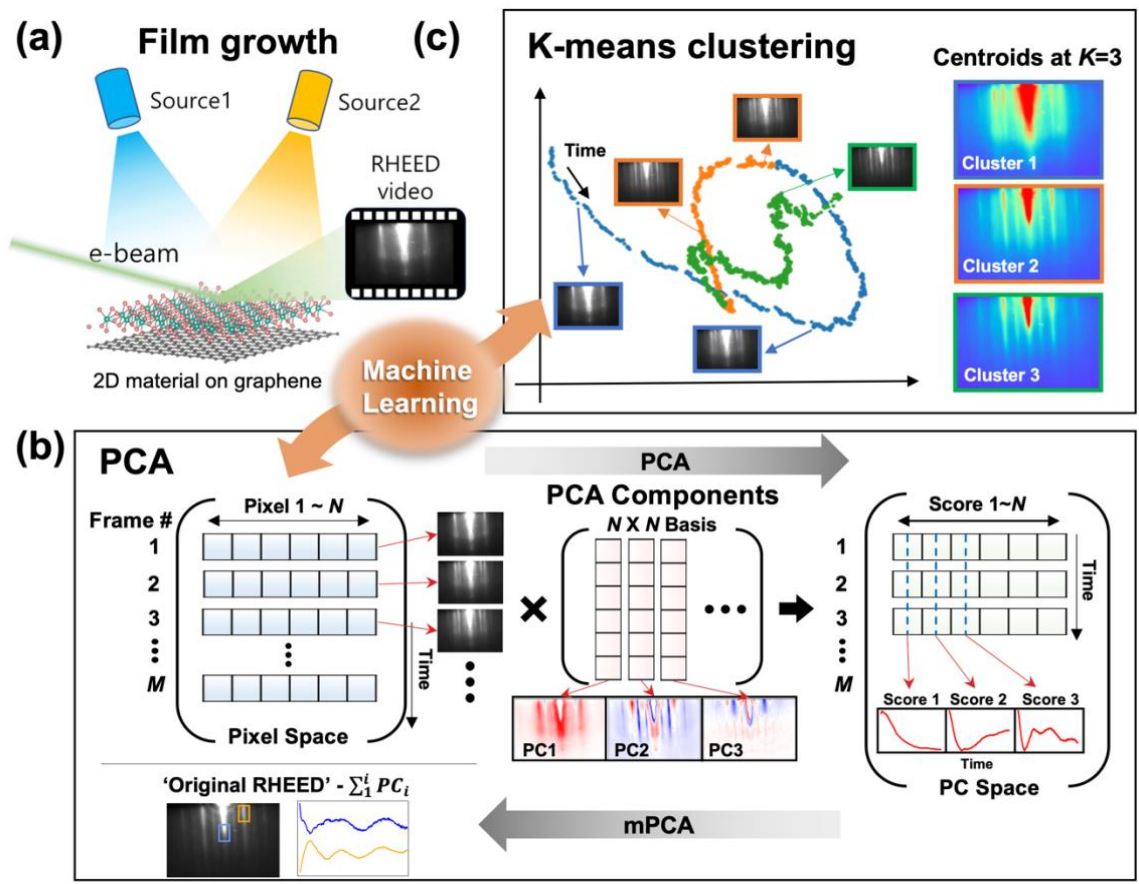

Figure 1. Overview of ML-assisted growth analysis. (a) Schematic of the growth of 2D layered thin films by MBE and acquisition of *in situ* RHEED video, (b,c) Processes of PCA and K-means clustering.

## 2. Results

We prepared ReSe$_2$ thin films, with varied thicknesses, on graphene substrates. Figure 2(a) shows the atomic structure of the distorted 1T (1T') ReSe$_2$. Figure 2(b-d) show the schematic models of the graphene substrate and ReSe$_2$ thin films with 0.3 and 3 unit cells (UC), respectively. We monitored the growth of ReSe$_2$ with *in situ* RHEED measurements and then compared the results with *ex situ* atomic force microscopy (AFM) data, as shown in Fig. 2(e-j). Initially, the bilayer graphene substrate was prepared with a sharp RHEED pattern (Fig. 2(e)) and a very flat surface with wide terraces (Fig. 2(h)). After 4 min of film growth, additional streaks of the ReSe$_2$ lattice emerged in the RHEED pattern, indicated by red arrows in Fig. 2(f). Further, small ReSe$_2$ islands were nucleated in the topography (Fig. 2(i)). After 62 min of deposition, the RHEED pattern of graphene completely disappeared, leaving only the ReSe$_2$ streaks, as shown in Fig. 2(g). The vertically elongated ReSe$_2$ streaks indicated a flat surface topography of the ReSe$_2$ thin film [9]. The in-plane lattice parameter of the ReSe$_2$ layer was estimated by comparing the RHEED streaks of graphene and ReSe$_2$. The calculated in-plane lattice parameter was 6.58 Å, which was consistent with the bulk value (6.60 Å($a_1$) and 6.71 Å($a_2$)) [26]. The corresponding ReSe$_2$ thin film showed a flat surface with a roughness of 0.23 nm (Fig. 2(j)), and its thickness was expected to be about 3UC.

The 3UC-thick ReSe$_2$ was characterized by Raman spectroscopy, as shown in Fig. 2(k). ReSe$_2$ exhibited diverse vibration modes in the range of 100–300 cm$^{-1}$, because the inversion symmetry is broken in 1T' ReSe$_2$. The peak positions were consistent with those of the ReSe$_2$ bulk and thick films, and the peak positions showed only a slight thickness dependence [27,28]. We also evaluated the layer thickness by high-angle annular dark field (HAADF) scanning transmission electron microscopy (STEM) analysis, as shown in Fig. 2(l). In this figure, three horizontal arrays of white dots are sandwiched with grey dots, as indicated by black arrows. Evidently, the top ReSe$_2$ layer shows a weaker signal, probably due to an incomplete coverage of the topmost layer. Additionally, we examined the stoichiometry of ReSe$_2$ by X-ray photoemission spectroscopy (XPS). We calculated the integrated peak areas of Re *4f* and Se *3d* and found that the Se/Re atomic ratio was approximately 2.01; this value was similar to the nominal stoichiometric ratio (see Supplementary Information, Fig. S1). These results confirm the successful growth of ReSe$_2$ thin films with controlled thicknesses, and indicate that the corresponding RHEED data can be analyzed by ML techniques.

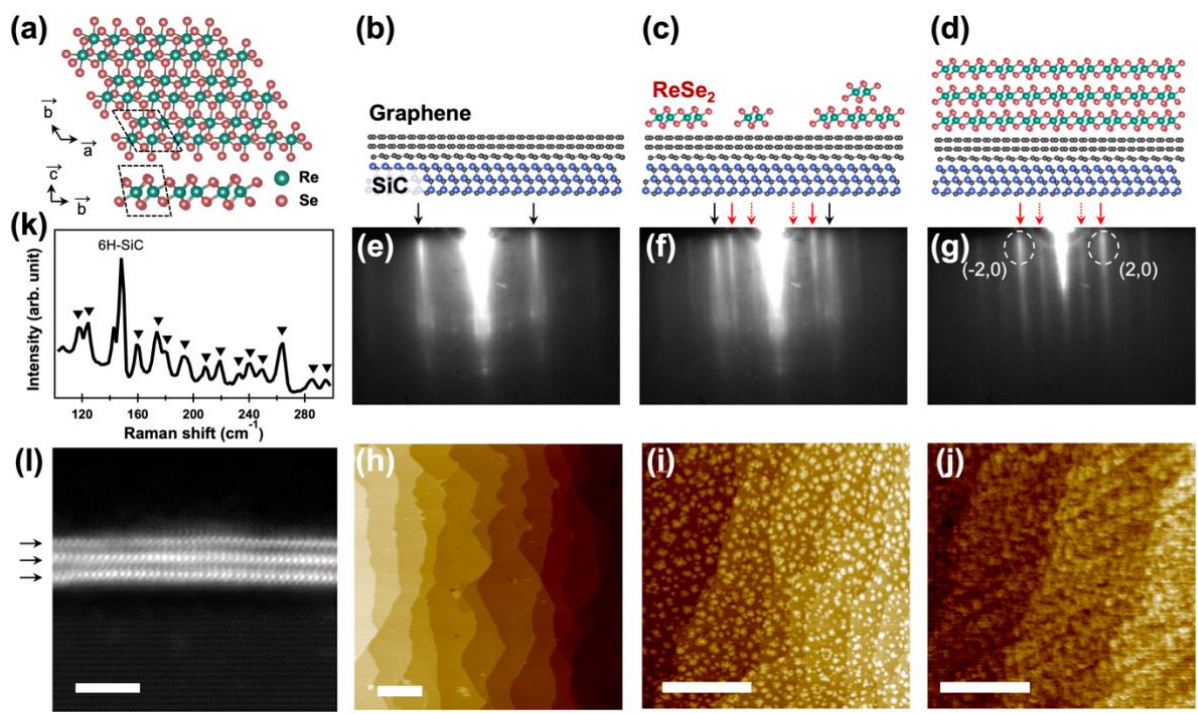

Figure 2. Growth and characterization of ReSe$_2$ thin films. (a) Crystal structures of 1T' ReSe$_2$. (b-d) Schematic models of the graphene substrate and ReSe$_2$ thin films with 0.3UC and 3UC. (e-g) RHEED images and (h-j) AFM images of the ReSe$_2$ thin film for different growth times (0, 4, and 62 min). The black and red arrows in the RHEED images indicate the bilayer graphene substrate and ReSe$_2$ diffraction streak, respectively. (k) Raman spectrum and (l) HAADF STEM image of the 3UC ReSe$_2$ film. Scale bars in the AFM and STEM images are 500 nm and 3 nm, respectively.

First, we analyzed the RHEED video of the ReSe$_2$ film by PCA. Figures 3(a,b) show the first six principal components (PCs) and their corresponding score values, which are similar to the concepts of eigenvectors and eigenvalues, respectively. The six components add up to 98.95 % of statistical variance in the dataset (see Supplementary Information, Fig. S2), implying most of the dataset can be represented by a few components and scores. Especially, PC1 has the most variation (91.98 %) in the RHEED video. The PC1 in Fig. 3(a) shows two major characteristics. First, the positive (red) area well matches the graphene pattern shown in Fig. 2(e). On the contrary, the negative (blue) area matches with the (2,0) and (-2,0) diffraction points of ReSe$_2$. The score 1, or the change in PC1 over time, decreases gradually and undergoes a sign change from positive to negative near the third dashed line in Fig. 3(b). This result implies that in the initial RHEED video, a gradually decreasing trend of the graphene signal is primarily observed. This signal trend is strikingly different from that of the oxide thin film, in which the in-plane lattice parameters are mostly nearly matched [19–21].

The second component, PC2 dominates the ReSe$_2$ streaks and minor diffraction points on the graphene and SiC substrates. The negative value of PC2 represents the epitaxial 2D growth of the ReSe$_2$ thin film, which is evidenced by the similar RHEED pattern of ReSe$_2$ in Fig. 2(g). The positive (red) region of PC2 includes the graphene diffraction streaks and several additional spots in the middle. Such spots are related to the buffer layer and SiC substrate beneath the graphene [29]. The initial decrease in score 2 (Fig. 3(b)) indicates that the substrate pattern disappears, and the ReSe$_2$ pattern begins to emerge, corresponding to the first dashed line.

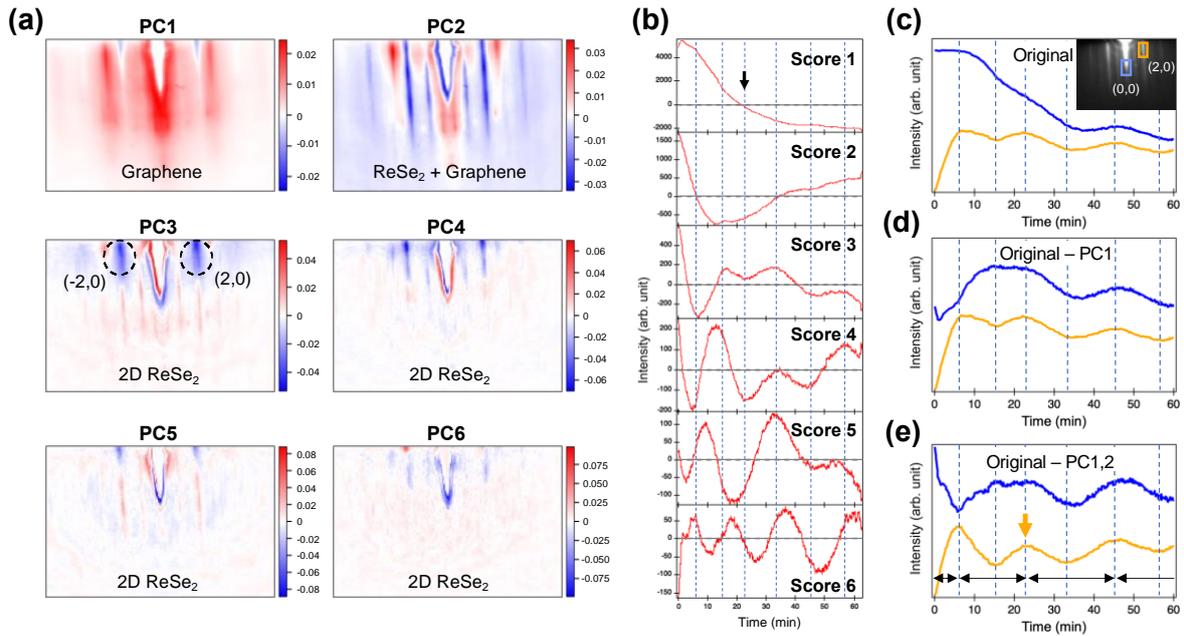

Figure 3. PCA results; (a) Six PCs of the RHEED video for the 3UC-thick ReSe$_2$ thin film and (b) the corresponding score plots. Component 1 (PC1) shows the diffraction signal of graphene, while component 2 (PC2) contains the signals of both the graphene and ReSe$_2$ layers. Component 3-6 (PC3-6) show the signal of only the 2D growth of ReSe$_2$ layer. (c-e) The intensity plots of the (c) original RHEED video and (d,e) modified RHEED video. Blue and orange lines denote the (0,0) and (2,0) diffraction streaks of the ReSe$_2$ thin film (shown in the inset), respectively.

Conversely, PC3–6 contain the (2,0) and (-2,0) diffraction signals of the 3UC ReSe$_2$ layers. The corresponding score 3–6 exhibit an oscillating behavior (Fig. 3(b)). In the MBE growth, the oscillating behaviors of specular or diffraction spots are used to estimate the film thickness and to analyze the growth modes [30]. In the layer-by-layer growth mode, the RHEED intensity is periodically modulated by the interference between the adjacent layers or the degree of diffused scattering, depending on the surface coverage [8]. The PCA results of other thicknesses (2UC, 4UC, and 5UC) also revealed that the oscillating character is observed when the PCs include the (2,0) and (-2,0) diffraction signals (see Supplementary Information, Fig. S3). Although the contribution of PC3–6 to the entire RHEED signal is <2 % (see Supplementary Information, Fig. S2), they contain physical meaning about the film thickness and its growth mode.

PCA is a versatile technique that allows us not only to decompose complex RHEED image sequences but also to selectively recombine the PCs and scores. However, further reconstruction of the selected components to extract the buried signal of interest has not yet been demonstrated. In the RHEED data of 3UC ReSe$_2$, we noticed that the strong signals of the graphene and substrate overshadowed the weak film intensities at the initial growth duration. In Fig. 3(c), the (0,0) peak intensity gradually declines and represents the graphene contribution, which is well correlated to score 1. To separate the weak ReSe$_2$ signal from the original video, we obtained the modified RHEED data (mPCA) by consecutively subtracting graphene-related components (PC1 or PC2) from the raw RHEED video, as described schematically in Fig. 1(b). Figs. 3(d,e) show the intensity plot of the (0,0) (blue lines) and (2,0) (orange lines) streaks obtained from the mPCA video sets. In Fig. 3(d), the subtraction of PC1 mainly changes the intensity plot within the initial period up to the third dashed line (23 min). This change indicates the signal transition from graphene to ReSe$_2$, consistent with the sign change in score 1 (indicated with an arrow in Fig. 3(b)). In Figs. 3(e), further subtractions of PC1 and PC2 result in stable oscillations for both blue and orange curves. Such oscillatory behaviors of the (0,0) and (2,0) streaks are likely linked to the layer-by-layer film growth, as mentioned before [8]. Interestingly, the orange curves show an additional period compared to the blue ones. Such discrepancy occurs in the initial duration when the strong graphene signal is overlapped with the ReSe$_2$ signal. In this duration, the blue curves show a dip and slow recovery up to 23 min, while the orange curves show a peak-dip-peak shape. The consistent oscillating behaviors of the blue and orange curves in Fig.3(e) provide accurate information about the film thickness such that the resulting film thickness of 3UC is consistent with the STEM data presented in Fig. 2(c). Accordingly, we added the vertical dashed lines in Figs. 3(b-e) and 4(a).

For comparison with the PCA results, we analyzed an identical RHEED dataset by K-means clustering. The K-means clustering method categorizes the sequence of the RHEED images into several clusters based on similarity without the need for complex mathematical transformations, and thus, determines the transition moments between distinct phases during the thin-film growth. It is worth noting that the relation between PCA and K-means algorithms is somewhat linked, as established well previously [19,31,32]. We employed a different number of clusters ($K$ = 2–6). Figures 4(a,b) show the time-dependent clustering for each $K$ value and the corresponding centroids. As $K$ is increased from 2 to 6, more divided sections appear for the initial growth time (i.e., < 35 min), implying that the major pattern change mostly occurs at the initial duration. The boundaries between the clusters show good alignment with the vertical dashed lines for $K$ = 5 and 6 (Fig. 4(a)). As shown in Fig. 4(c), the cost function (i.e., the accumulated differences between the clusters and the original data) is used to determine the valid number of clusters, and the appropriate $K$ is near the saturation point of the curve [21]. The cost function is saturated when $K$ > 4. To investigate the evolution of the centroids in detail, we plotted the difference between the adjacent centroids ($\Delta C_{i(i+1)}$) as shown in Fig. 4(d) by subtracting a former centroid ($C_i$) from a latter one ($C_{i+1}$) for $K$ = 6. Here, the positive (red) and negative (blue) regions represent the emerging and disappearing characteristics in the RHEED patterns, respectively. A distinct feature of $\Delta C_{12}$ is the emerging ReSe$_2$ streak signal (indicated with red arrows), which corresponds to the emerging ReSe$_2$ signal in the PCA. The graphene signal (black arrows) shows a gradually disappearing trend up to $\Delta C_{45}$ (23 min). This boundary corresponds to the third dashed line, at which the graphene signal nearly disappears as score 1 becomes negative in the PCA (Fig. 2(b)). After the graphene signal disappears, $\Delta C_{56}$ mostly shows the intensity variations in the ReSe$_2$ streaks, implying a homoepitaxial growth regime. Therefore, the results obtained by K-means clustering with $K > 4$ were consistent with those of the PCA.

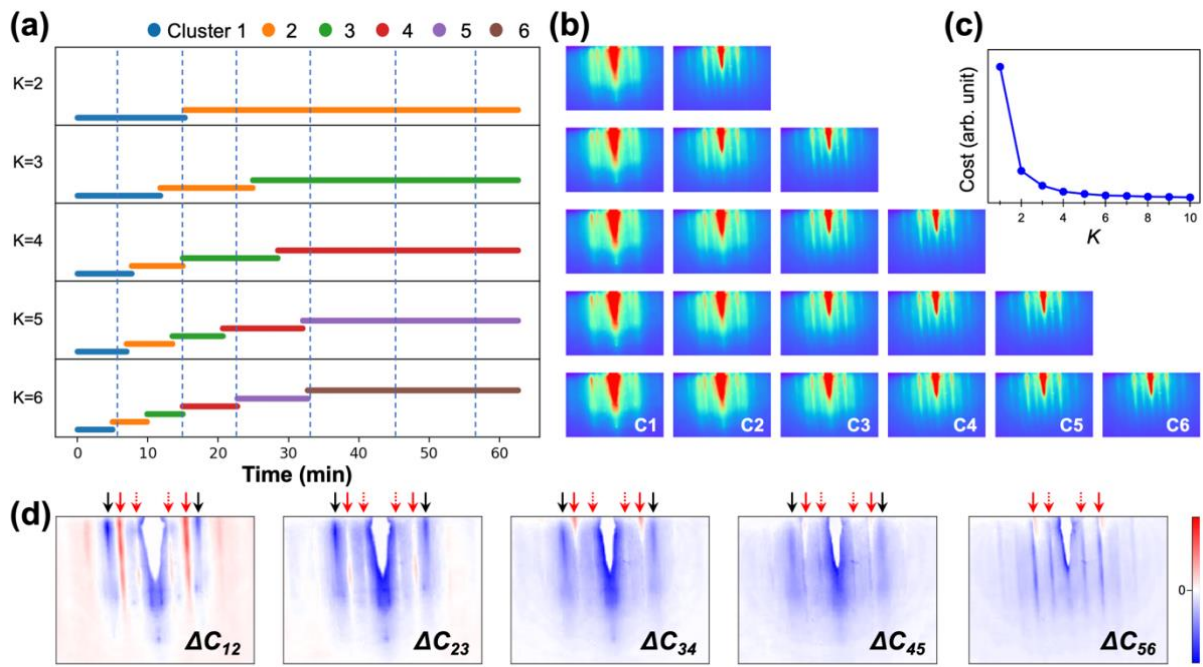

Figure 4. K-means clustering analysis of the RHEED video of the 3UC ReSe$_2$. (a) Clusters with number of clusters ($K$ = 2–6) and (b) their corresponding centroids. (c) Cost function as a function of $K$. (d) Difference between the adjacent centroids for $K$ = 6.

## 3. Discussion

The stable oscillations of RHEED diffraction streaks in Fig. 3(e) indicate that the ReSe$_2$ film growth nearly follows the layer-by-layer growth mode. The two oscillation peaks are observed until the RHEED signal of graphene disappears, as shown by an orange arrow (~23 min) in Fig. 3(e). This observation corresponds to the sign reversal moment of score 1 (Fig. 3(b)). The two oscillation peaks imply that the small portion of bilayer $ReSe_2$ domains are formed, before the graphene surface is completely covered, at the given growth condition. Such a phenomenon was observed in our previous scanning tunneling microscopy-based study, in which we had observed the partial formation of bilayer $ReSe_2$ islands when the graphene surface was incompletely buried [23]. These results of $ReSe_2$ growth behavior suggest some deviation from the layer-by-layer growth mode to the Stranski-Krastanov growth mode,

Moreover, the first oscillation period of the (2,0) streak is approximately half of the following oscillation periods (black arrows in Fig. 3(e)). The shrinking of the first oscillation indicates that either the growth rate in the first layer accelerated or that in the following layers decelerated. Abrupt changes in the RHEED oscillation occur in the case of $SrRuO_3$ growth on $SrTiO_3$ (001) surfaces [33], and the first oscillation period is two times longer than the following periods. Koster et al. concluded that $RuO_x$ re-evaporates, and the growth rate of the first $SrRuO_3$ layer drops to nearly half of its initial value [34]. This decrease in the growth rate implies that the growth dynamics of the first layer are largely dependent on the surface energy of the substrates in the case of complex oxides and chalcogenides [33–36]. In our case, the film growth process can be divided into two situations: $ReSe_2$ layer on graphene surface (heteroepitaxy) and $ReSe_2$ layer on $ReSe_2$ surface (homoepitaxy). Assuming that the number of atoms that are deposited is kept same during the film growth, the different surface energies of graphene and $ReSe_2$ are expected to lead to a faster growth of the first $ReSe_2$ layer when it is grown on a graphene. The shortening of the first RHEED oscillations are consistently observed when several $ReSe_2$ films are repeated (see Supplementary Information, Fig. S4). Since different substrate surface states have also shown alteration of growth modes of TMDC thin films [35,36], further analysis of the initial RHEED analysis for different substrates and thin film materials would be beneficial to investigate the correlation between the surface energy and growth modes [37].

We applied comprehensive ML analyses, such as PCA and K-means clustering, to understand the growth mechanism of an $ReSe_2$ thin film on graphene, which is a model van der Waals heteroepitaxial system. In case of the oxide film growth, the previous ML analyses of RHEED have reported the growth modes and the implication of PCs because the RHEED patterns maintain similar shapes and sizes from the substrates to the films. However, TMDC thin films have been successfully grown on substrates with largely mismatched lattices, such as graphene and sapphire, because of the weak van der Waals bonding at the interfaces [1]. The low-dimensionality characteristic of the TMDCs also gives rise to unique layer-dependent quantum phenomena. Thus, precise prediction of the film thickness is crucial for the initial growth. The dominant substrate signal in the RHEED pattern hinders the analysis of the initial growth mechanism of a thin film. Our ML analysis focused on separating PCs corresponding to the substrates and the films by utilizing PCA with statistical significance. This ML analysis is beneficial for analyzing the growth dynamics and layer thicknesses for ultrathin van der Waals thin films, and the corresponding results are consistent with those of the K-means clustering method. Our results suggest that the ML-assisted RHEED analysis could be developed into an automatic validation method for investigating ultrathin 2D materials films, and it is complementary to other surface analysis tools [7,38,39]. Furthermore, this method can be applied to analyze the thin-film growth of other 2D materials, such as 2D chalcogenides, 2D MXenes, 2D oxides, and hexagonal boron nitrides [40–43].

## 4. Conclusions

In summary, we conducted an ML-assisted *in situ* RHEED analysis to understand the epitaxial growth of ReSe$_2$ thin films, with different thicknesses, on graphene. Using PCA, we can separate the *in situ* RHEED dataset into newly defined PCs and their scores based on their statistical significance. We observed the growth dynamics of the ReSe$_2$ thin film by subtracting the graphene substrate contribution. We confirmed that the time evolution of the *K*-means clusters for $K > 4$ was consistent with the PCA result. Therefore, these results indicate the feasibility of applying ML techniques to analyze the epitaxial growth of 2D layered materials and suggest that such techniques can accelerate the development of automated film growth processes.

## 5. Experimental section

### 5.1 Film growth

ReSe$_2$ thin films were grown on an epitaxial graphene bilayer, which was fabricated on a (0001) 6H-SiC substrate, using a home-built MBE system in ultrahigh vacuum (base pressure: $1.0 \times 10^{-9}$ torr). For the growth of the bilayer graphene on the SiC substrate, the substrate was outgassed at 650 °C for a few hours, and the substrates were subsequently annealed at 1300 °C for 6 min, as verified by the RHEED image shown in Fig. 2(f). High-purity Re (99.8 %) and Se (99.999 %) were used for the ReSe$_2$ thin-film growth. We synthesized the ReSe$_2$ thin film by co-evaporating Re and Se using an electron-beam evaporator and a Knudsen cell, respectively, while monitoring the film surface by the *in situ* RHEED, as shown in Fig. 1(a). The substrate was maintained at 300 °C during the deposition [44].

### 5.2 Characterization

The Raman spectroscopic measurements were performed using a 532 nm excitation laser source with a fixed power (30 mW) and fixed acquisition time (60 s) at room temperature.

Scattered light from the samples was analyzed using a single-grating monochromator with a focal length of 50 cm, and was detected by a liquid-nitrogen-cooled charge-coupled-device detector (LabRAM HR Evolution, HORIBA). AFM was performed to investigate the surface morphology under atmospheric conditions after the deposition (XE-100, Park system), and the samples were scanned in the non-contact mode using an NSC18/Pt tip. The XPS measurements were carried out to examine the stoichiometry of the films (NEXSA, Thermo Fisher Scientific). For the STEM analysis, cross-sectional specimens were fabricated using the focused ion beam technique (Helios Nanolab 450, ThermoFisher Scientific). The HAADF STEM images were obtained using a double Cs-corrected FEI Titan G2 60-300 microscope with an accelerating voltage of 200 kV.

**5.3 ML method**

All the ML analyses were carried out using python version 3.8.12 (the code and model are publicly available [44]). For the PCA, first, we converted the RHEED video into a 2D array, namely $X$, which was an $M \times N$ matrix, where $M$ and $N$ represented the number of frames and pixels, respectively. We captured RHEED video at a rate of one frame/second so that each row of the matrix represented an RHEED image at a particular time as shown in Fig. 1(b) (blue-shaded boxes). For the PCA, the dataset was converted into a linearly superposed set with component weights and orthogonal basis consisting of eigenvectors. The basis matrix (red-shaded boxes) formed an $N \times N$ matrix, and the column vectors indicated the individual PCs. In this newly defined matrix (green-shaded boxes in the PC space), the components were determined by the production of $X$ and the basis matrix. The row vectors represented the RHEED images arranged in a descending order of eigenvalues ('Score'), whereas the column vectors represented the time-dependent behavior of each score. We proposed a reconstruction process, namely the mPCA, in which the frames of the PC space were merged while eliminating some of the selected PCs, i.e. "Original RHEED" - $\sum_{i=1}^{n} PC_i$, for eliminating the substrate contributions. Then, we extracted the time dependences of the selected diffraction peak intensities, as shown in the left bottom of Fig. 1(b).

Next, we carried out the K-means clustering analysis by using 20 PCs to reduce the dimension of the original dataset for faster computing. We split the RHEED image series into $K$ clusters, in which each image was classified to the cluster with the nearest mean ("centroid"). First, we randomly selected the $K$ images, as the initial selections of the centroids, from the whole dataset. Then, we allocated each RHEED image to the nearest centroid. The old centroids were replaced by the mean images constituting the corresponding clusters. This replacement was iteratively repeated until the centroids stopped changing.

**Abbreviations**

RHEED: Reflective high-energy electron diffraction, MBE: Molecular beam epitaxy, ML: Machine learning, PC: Principal components, PCA: Principal component analysis, mPCA: Modified principal component analysis, 2D: Two-dimensional, TMDC: Transition metal dichalcogenide, 1T': Distorted 1T, UC: Unit cell, AFM: Atomic force microscopy, HAADF: High-angle annular dark field, STEM: Scanning transmission electron microscopy, XPS: X-ray photoemission spectroscopy.

**<u>Declarations</u>**

**Acknowledgments**

Authors thank Yea-Lee Lee and Seunghun Jang for the constructive discussions.

**Availability of data and materials**

The machine learning codes are available in GitHub at https://github.com/youngjunchang/RHEED_2D_ML, Ref. 45, and the RHEED videos are available in our web-based platform, *2D Materials* at http://2dmat.chemdx.org/data_uos, Ref 46, with permission from the corresponding authors upon reasonable request.

**Competing interests**

The authors declare no competing financial interest.

**Authors' information**

[1]Department of Physics, University of Seoul, Seoul, 02504, Republic of Korea.

[2]Department of Smart Cities, University of Seoul, Seoul, 02504, Republic of Korea.

[3]Department of Energy Science, Sungkyunkwan University (SKKU), Suwon, 16419, Republic of Korea.

[4]Graduate School of Semiconductor Materials and Devices Engineering, Ulsan National Institute of Science and Technology (UNIST), Ulsan, 44919, Republic of Korea.

[5]Advanced Light Source (ALS), E. O. Lawrence Berkeley National Laboratory, Berkeley, California 94720, USA.

**Funding**

This work was supported by the National Research Foundation of Korea (NRF) grants funded by the Korea Government (NRF-2020R1A2C200373211, 2021R1A6A3A14040322, and 2022R1A2C2011109) and [Innovative Talent Education Program for Smart City] by MOLIT. This research has been performed as a cooperation project of "Basic project (referring to projects performed with the budget directly contributed by the government to achieve the purposes of establishment of government-funded research institutes)" and supported by the Korea Research Institute of Chemical Technology (KRICT) (S12151-10-06).

**Authors' contributions**

HJK and YJC designed the experiments and performed the analyses. HJK, YGK, TGR, and BKC performed the sample preparation and characterizations. M-HJ, Y-MK and HYJ carried out the scanning transmission electron microscopy analysis. HJK and MC carried out the machine-learning analysis. HJK, MC and YJC analyzed the results and prepared the manuscript. All authors read and approved the final manuscript.

**References**

1. T. H. Choudhury, X. Zhang, Z. Y. Al Balushi, M. Chubarov, and J. M. Redwing, Annu. Rev. Mater. Res. **50**, 155 (2020).

2. S. M. George, Chem. Rev. **110**, 111 (2010).

3. Y.-L. Huang, H.-J. Liu, C.-H. Ma, P. Yu, Y.-H. Chu, and J.-C. Yang, Chinese J. Phys. **60**, 481 (2019).

4. P. Ranjan, S. Gaur, H. Yadav, A. B. Urgunde, V. Singh, A. Patel, K. Vishwakarma, D. Kalirawana, R. Gupta, and P. Kumar, Nano Converg. **9**, 26 (2022).

5. T. Orvis, M. Surendran, Y. Liu, A. Cunniff, and J. Ravichandran, J. Vac. Sci. Technol. A **37**, 61401 (2019).

6. J. H. Gruenewald, J. Nichols, and S. S. A. Seo, Rev. Sci. Instrum. **84**, 43902 (2013).

7. Y. Li, F. Wrobel, Y. Cheng, X. Yan, H. Cao, Z. Zhang, A. Bhattacharya, J. Sun, H. Hong, H. Wang, Y. Liu, H. Zhou, and D. D. Fong, ACS Appl. Mater. Interfaces **14**, 16928 (2022).

8. A. Ichimiya, P. I. Cohen, and P. I. Cohen, *Reflection High-Energy Electron Diffraction* (Cambridge University Press, 2004).

9. G. Liang, L. Cheng, J. Zha, H. Cao, J. Zhang, Q. Liu, M. Bao, J. Liu, and X. Zhai, Nano Res. **15**, 1654 (2022).

10. N. J. C. Ingle, A. Yuskauskas, R. Wicks, M. Paul, and S. Leung, J. Phys. D. Appl. Phys.

**43**, 133001 (2010).

11. P. K. Larsen and G. Meyer-Ehmsen, Surf. Sci. **240**, 168 (1990).

12. J. Jo, Y. Tchoe, G.-C. Yi, and M. Kim, Sci. Rep. **8**, 1694 (2018).

13. R. Shimizu, S. Kobayashi, Y. Watanabe, Y. Ando, and T. Hitosugi, APL Mater. **8**, 111110 (2020).

14. B. P. MacLeod, F. G. L. Parlane, T. D. Morrissey, F. Häse, L. M. Roch, K. E. Dettelbach, R. Moreira, L. P. E. Yunker, M. B. Rooney, J. R. Deeth, V. Lai, G. J. Ng, H. Situ, R. H. Zhang, M. S. Elliott, T. H. Haley, D. J. Dvorak, A. Aspuru-Guzik, J. E. Hein, and C. P. Berlinguette, Sci. Adv. **6**, eaaz8867 (2022).

15. G. S. Na, S. Jang, and H. Chang, Npj Comput. Mater. **7**, 106 (2021).

16. Y.-L. Lee, H. Lee, T. Kim, S. Byun, Y. K. Lee, S. Jang, I. Chung, H. Chang, and J. Im, J. Am. Chem. Soc. **144**, 13748 (2022).

17. J. Kwoen and Y. Arakawa, Cryst. Growth Des. **20**, 5289 (2020).

18. H. Liang, V. Stanev, A. G. Kusne, Y. Tsukahara, K. Ito, R. Takahashi, M. Lippmaa, and I. Takeuchi, Phys. Rev. Mater. **6**, 63805 (2022).

19. R. K. Vasudevan, A. Tselev, A. P. Baddorf, and S. V Kalinin, ACS Nano **8**, 10899 (2014).

20. K. Gliebe and A. Sehirlioglu, J. Appl. Phys. **130**, 125301 (2021).

21. S. R. Provence, S. Thapa, R. Paudel, T. K. Truttmann, A. Prakash, B. Jalan, and R. B. Comes, Phys. Rev. Mater. **4**, 83807 (2020).

22. S.-K. Mo, Nano Converg. **4**, 6 (2017).

23. T. Thi Ly, Y.-J. Lee, B. Ki Choi, H. Lee, H. Jin Kim, G. Duvjir, N. Huu Lam, K. Jang, K. Palotás, Y. Jun Chang, A. Soon, and J. Kim, Appl. Surf. Sci. **579**, 152187 (2022).

24. Y.-Y. Chang, H. N. Han, and M. Kim, Appl. Microsc. **49**, 10 (2019).

25. X. Cong, X.-L. Liu, M.-L. Lin, and P.-H. Tan, NPJ 2D Mater. Appl. **4**, 13 (2020).

26. H.-J. Lamfers, A. Meetsma, G. A. Wiegers, and J. L. de Boer, J. Alloys Compd. **241**, 34 (1996).

27. Y. Choi, K. Kim, S. Y. Lim, J. Kim, J. M. Park, J. H. Kim, Z. Lee, and H. Cheong, Nanoscale Horizons **5**, 308 (2020).

28. D. Wolverson, S. Crampin, A. S. Kazemi, A. Ilie, and S. J. Bending, ACS Nano **8**, 11154 (2014).

29. I. S. Kotousova, S. P. Lebedev, A. A. Lebedev, and P. V Bulat, Phys. Solid State **61**, 1940 (2019).

30. M. Nakano, Y. Wang, Y. Kashiwabara, H. Matsuoka, and Y. Iwasa, Nano Lett. **17**, 5595 (2017).

31. C. Ding and X. He, in *Proc. Twenty-First Int. Conf. Mach. Learn.* (ACM Press, 2004), p.

29.

32. P. Drineas, A. Frieze, R. Kannan, S. Vempala, and V. Vinay, Mach. Learn. **56**, 9 (2004).

33. Y. J. Chang and S. Phark, ACS Nano **10**, 5383 (2016).

34. G. Koster, L. Klein, W. Siemons, G. Rijnders, J. S. Dodge, C.-B. Eom, D. H. A. Blank, and M. R. Beasley, Rev. Mod. Phys. **84**, 253 (2012).

35. W. Mortelmans, A. Nalin Mehta, Y. Balaji, S. Sergeant, R. Meng, M. Houssa, S. De Gendt, M. Heyns, and C. Merckling, ACS Appl. Mater. Interfaces **12**, 27508 (2020).

36. W. Wan, L. Zhan, T.-M. Shih, Z. Zhu, J. Lu, J. Huang, Y. Zhang, H. Huang, X. Zhang, and W. Cai, Nanotechnology **31**, 35601 (2019).

37. A. Rajan, K. Underwood, F. Mazzola, and P. D. C. King, Phys. Rev. Mater. **4**, 14003 (2020).

38. H. K. Yoo, D. Schwarz, S. Ulstrup, W. Kim, C. Jozwiak, A. Bostwick, T. W. Noh, E. Rotenberg, and Y. J. Chang, J. Korean Phys. Soc. **80**, 1042 (2022).

39. R. Kim, B. K. Choi, K. J. Lee, H. J. Kim, H. H. Lee, T. G. Rhee, Y. G. Khim, Y. J. Chang, and S. H. Chang, Curr. Appl. Phys. **46**, 70 (2023).

40. B. C. Wyatt, S. K. Nemani, and B. Anasori, Nano Converg. **8**, 16 (2021).

41. A. Iqbal, J. Hong, T. Y. Ko, and C. M. Koo, Nano Converg. **8**, 9 (2021).

42. H. He, Z. Yang, Y. Xu, A. T. Smith, G. Yang, and L. Sun, Nano Converg. **7**, 32 (2020).

43. R. Page, J. Casamento, Y. Cho, S. Rouvimov, H. G. Xing, and D. Jena, Phys. Rev. Mater. **3**, 64001 (2019).

44. B. K. Choi, S. Ulstrup, S. M. Gunasekera, J. Kim, S. Y. Lim, L. Moreschini, J. S. Oh, S.-H. Chun, C. Jozwiak, A. Bostwick, E. Rotenberg, H. Cheong, I.-W. Lyo, M. Mucha-Kruczynski, and Y. J. Chang, ACS Nano **14**, 7880 (2020).

45. https://github.com/youngjunchang/RHEED_2D_ML.

46. https://2dmat.chemdx.org/data_uos.